\documentclass[a4paper,aps,prl,twocolumn,floatfix]{revtex4}
\usepackage{epsfig}
\usepackage{amsmath}
\usepackage{graphics}
\usepackage{graphicx}

\def\beq{\begin{equation}}
\def\eeq{\end{equation}}
\def\bea{\begin{eqnarray}}
\def\eea{\end{eqnarray}}

\def\u1{\widehat{U(1)}}
\def\su2{\widehat{su(2)}_1}

\def\uqan{U_q(\widehat{sl(n)})}

\def\uqa{U_q(\widehat{sl(2)})}

\def\a{\alpha}
\def\b{\beta}
\def\g{\gamma}

\def\l{\lambda}

\def\de{\partial}

\def\nn{\nonumber}

\begin{document}
\title{Mott transition and integrable lattice models in two dimensions}

\author{Federico L. Bottesi}
\author{Guillermo R. Zemba}
\affiliation{ Facultad de Ingenier\'ia  Pontificia Universidad Cat\'olica Argentina, Av Alicia Moreau de Justo 1500, 1428, Buenos Aires, Argentina\\} 
\affiliation{Departamento de F\'{\i}sica, Laboratorio Tandar,
         Comisi\'{o}n Nacional de Energ\'{\i}a At\'{o}mica, Avenida del
         Libertador 8250, 1429 Buenos Aires, Argentina}

\date{\today}

\begin{abstract}
We  describe the two-dimensional Mott transition in a Hubbard-like model with nearest neighbors interactions
based on a recent solution to the Zamolodchikov tetrahedron equation, which extends the notion 
of integrability to two-dimensional lattice systems.  At the Mott transition, we find that  
the system is in a $d$-density wave or staggered flux phase that can be described by a 
double Chern Simons effective theory with symmetry  $\su2 \otimes \su2$.
The Mott transition is of topological nature, characterized by the emergence of vortices in 
antiferromagnetic arrays interacting strongly with the electric charges
and an electric-magnetic duality.
We also consider the effect of small doping on this theory and show that it  
leads to a quantum gas-liquid coexistence phase, which belongs to the Ising universality class
and which is consistent  
with several experimental observations.  
\end{abstract}

\maketitle

In spite of substantial advances in our theoretical understanding of strongly correlated electron systems, several problems
still continue to provide estimulating challenges. One of the most interesting among these is the Mott transition, or metal-insulator transition driven by correlations. As early as in 1939, Mott argued that if the electron density in a metallic system was lowered enough, the Coulomb repulsion would dominate over the kinetic energy so that the system would 
undergo a transition to an insulating regime \cite{Mott}. From the experimental point of view, there exist several systems which display a Mott-type transition, such as vanadium oxide $V_2O_3$, several organic conductors, some doped semiconductors, and even underdoped high $T_C$ superconductors. 
Moreover, coexistence between phases of different densities has been observed in several experiments \cite{Algun-Exp}. From the theoretical point of view, finding solutions to even the simplest models (such as the Hubbard model) is difficult, given the failure of  perturbative approaches due to the narrow differences separating the localized regime of the electrons in the insulating phase and the itinerant one in the conducting state. 

As of today, there exist two basic approaches for studying this transition: one is the dynamical mean field theory (DMFT) method\cite{Kotliar}, valid in the limit of infinite dimensions (or infinite coordination number), which maps the Hubbard model onto the impurity Anderson model, with the addition of a self-consistency  condition. This framework neglects spatial correlations while retaining the on-site quantum ones. The second approach consists in finding
analytic expressions for physical observables in integrable models exhibiting the Mott behavior, for example, by using the Bethe Ansatz or the bosonization methods \cite{Lieb} \cite{Giamarchi} \cite{Shankar}.
However, the main restriction of these models is that they are formulated in one spatial dimension, unlike most system of experimental interest. The goal of the present article is to extend this approach to a two-dimensional (integrable) lattice model that exhibits the Mott transition and writ down an Efective fiel Theory to futher analyze the behavior af the model.

Let us start by consider a system of spinless fermions on a square (two-dimensional) lattice
with hamiltonian:
\bea
H&=&-\frac{t}{2}\sum_{i,\mu} [\psi^\dagger(i+e_\mu) e^{i A_\mu} \psi(i)+h.c.] \nn\\ 
&& +U\sum_{i,\mu} \rho(i)\rho(i+e_\mu)\ , \label{Model-Ferm-2d}\eea 
where $i$ labels the lattice sites and $e_\mu$ are unit lattice vectors, $t$ is the hopping parameter, $U$ is the (constant) Coulomb potential, $\rho(i)$ is the normal ordered charge density with respect to the half-filling ground state,  $\rho(i)=:\psi^{\dagger}_i\psi_i: -1/2$ and $A_\mu $ is the abelian statistical gauge field, which after imposing Gauss' law constraint reads
\beq
A_\mu(i)=\sum_k[\Theta(k,i)-\Theta(k,i+e_\mu)]\psi^\dagger_k \psi_k\ , \eeq
where $\Theta(k,i)$ is the angle between the chosen direction $i$ and an arbitrary one $k$ on the lattice. Note that, 
for the one-dimensional case, the gauge field is irrelevant, in agreement with the fact that quantum statistics in one spatial dimension is arbitrary  \cite{Fradkin-Book} and does not involve any physical gauge field.
Using the two-dimensional Jordan-Wigner transformation\cite{Tsvelik}\cite{Fradkin-Book} :
\bea
&& S^+_j=\psi^\dagger_j U_{2d}(j) \nn\\
&& S^-_j=U_{2d}(j)\psi _j  \nn  \\
&& S^z=\psi^{\dagger} _j \psi_j -\frac{1} {2}\ ,\eea
where $U_{2d}(j)=\exp{[i\sum_{k\neq j}\Theta(k,j) \psi ^\dagger_k\psi_k]}$,the hamiltonian (\ref{Model-Ferm-2d})becomes that of a $XXZ$ Heisenberg model
\beq
H_{XXZ}=\sum_{\langle i,j \rangle} [ -(S^x_jS^x _j + S^y_i S^y_j)+\Delta S^z_jS^z_j ] \ , \label{hxxz}
\eeq 
where we have rescaled the terms such that  $\Delta=U/t$. Following  \cite{Jakel-Maillard} we define an {\it interaction star} as the set of points where the spins entering in an elementary interaction are localized, {\it i.e.}, the central site and their nearest-neighbors in the $XXZ$ model. The $n$-th interaction star has an energy  $E_{XXZ}([\sigma]_n)$  which depends on the spin configuration in the star and on the local Boltzmann weights 
$W([\sigma]_n)$ . 
Therefore, the partition function takes the form $Z=\sum_{\sigma}\prod_n W([\sigma]_n)$, where the sum is taken over all possible configurations of the entire lattice. 

\begin{figure}[h!]
\centering
\resizebox{7.cm}{!}{ 
\includegraphics[clip]{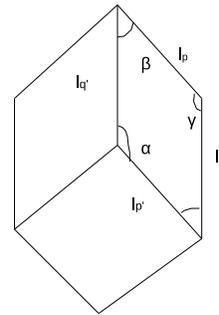}}
 \caption{\small{Diagram of a hexahedron in which the angles $(\alpha, \beta, \gamma ) $ and side lenghts $(l_p, l'_p , l_q , l'_q)$ }
for one face are indicated.}
\label{Fig1}
\end{figure}

In two dimensional quantum systems  and three dimensional statistical models  the integrability is guaranteed by the existence of a set of mutual commuting layer-to-layer transfer matrices $T_{mn}(\l,\mu)$, which is  tantamount to the existence of solutions of the so-called Zamolodchikov's tetrahedron equation(TE) \cite{Zamolodchikov}\cite{Bazhanov-1} : 
\beq
R_{abc}R_{ade}R_{bdf}R_{cef}=R_{cef}R_{bdf}R_{adc}R_{abc} \ ,\label{tetraeq}\eeq
where the operators $R_{ijk}$ define the mapping  $ R_{ijk}: V_i \otimes V_j \otimes V_k \rightarrow V_i \otimes V_j \otimes V_k $, and $V_n$ 
is the spin one-half representation space , such that their matrix elements are the Boltzmann weights of the vertex $R_{ijk}=W([\sigma]_{ijk})$( the indices $i,j,k$ label the interaction star). These can be rewritten as the $LLLR-RLLL$ operator conditions, which express the associativity of the Zamolodchikov algebra:
\beq
L_{12,a}L_{13,b}L_{23,c}R_{abc}=R_{abc}L_{23c}L_{13,c}L_{13,b}\ , \label{LLLR-RLLL}\eeq
where, for example, the operator $L_{12a}$ acts on $V_1 \otimes V_2\otimes F_a $,  $V_1$, $V_2$ are the auxiliary spaces and $F_a$ is the quantum space. If $F_a$ is the representation space of some algebra $\cal{A}$, it is possible to interpret the operators $L_{ij,a}$ as operator-valued matrices acting on $V_1 \otimes V_2$, and depending `parametrically' on the generators of the algebra $\cal{A}$ denoted by $v_a$ and, possibility, on some $c$-numbers denoted by $s_a$: $L_{12a}=L_{12}(v_a,s_a)$ . In this case, the equation (\ref{LLLR-RLLL}) can be expressed as a `local Yang Baxter' equation:
\bea
&& L_{12}({\bf v}_a,s_a) L_{13}({\bf v} _b,s_b) L_{23}({\bf v} _c,s_c)= \nn \\
                &&       \qquad L_{23}({\bf v'}_c,s_c)  L_{13}({\bf v'}_b,s_b)  L_{1a}({\bf v'}_a,s_a) \ . \label{Local-Yang-Baxter}\eea
The tetrahedron equation (\ref{tetraeq}) is highly non-trivial to solve, but recently a new solution to it has been found in \cite{Bazhanov-1}.
The solution is associated  to the finite-dimensional highest-weight representations of the quantum affine algebras $\uqa$, displaying the three-dimensional structure of quantum groups. It may be understood as a quantization of the spatial fluctuations of geometrical extended objects, and we shall see that in our case that  these may be reinterpreted as (discrete) charge density waves. For completeness, we now briefly review the new solution (for details see \cite{Bazhanov-1}).
The solution is inspired in the geometry of transformations applied to an hexahedron (see fig.(\ref{Fig1}): there are three independent angles on each face and nine angles to fix the spatial orientation of the hexahedron. Therefore, nine independent angles are needed to
specify it. Let us consider the mapping,
\beq
{\it R}_{123}: [\a_j,\b_j ,\gamma_j] \rightarrow [\a '_j,\b '_j ,\gamma '_j]\ ,\label{mapeofuncional-1} \eeq 
where $\a$, $\b$ and $\gamma$ are the angles of the $j$-th face, and the primed variables refer to the opposite faces.
For each quadrilateral face, say the $1'$, the relationship between opposite sides  is given by $(l'_p,l'_q)^t=X({\cal A}_1) (l_p,l_q)^t$, where $X({\cal A}_1)$ is a matrix acting non trivialy on the face $1'$ , that depends on the angles on that face (${\cal A}_1=(\a_1,\b_1,\g_1$)), and which for a circular lattice reads:
\[X ({\cal A}_1)  =  \left [ \begin{array}{ccc}
{k_1} & {a^*_1} & {0}           \\
{-a_1} & {k_1} &{ 0}  \\
 {0} & {0} & {1}\  \\
  \end{array}\right]\]
where  $k_1=\cos \a _1 \sin \b _1$, $a=\cos \a_1 \sin(\a _1+\b_1)$ and $a_1^*=\cos \a_1 \sin (\a_1-\b_1)$.
In the general case, considering three faces , we have:
\beq
(l'_p,l'_q,l'_r)^t=X_{pq}({\cal A}_1)X_{qr}({\cal A}_2)X_{rs}({\cal A}_3)(l_p,l_q,l_r)^t\ .\label{longitudes}\eeq
The same result (\ref{longitudes}) is obtained by using  the opposite faces with angles (${\cal A}'_1$,${\cal A}'_2$,${\cal A}'_3$). Therefore , it is easy to see that there exists a functional mapping given by $ X_{pq}({\cal A}_1)X_{qr}({\cal A}_2)X_{rs}({\cal A}_3)={\it R}_{123} (X_{rs}({\cal A}_3) X_{qr}({\cal A}_2)X_{pq}({\cal A}_3))$.  This relation could be considered as a 'gauge symmmetry'.
It can be shown that the mapping  ${\it R}_{123}$ satisfies the functional tetrahedron equation ($FTE$), a 'classical version' of the
tetrahedron equation:
\bea
&& k'_2a'^*_1=k_3a^*_1-k_1a^*_2a_3 \qquad  k'_2a'_1=k_3a_1-k_1a^*_2a^*_3 \nonumber\\    
&& a'^*_2=a^*_2a^*_3 +k_1k_3a^*_2 \qquad \quad   a_2'=a_2a_3 +k_1k_3a_2 \nonumber\\ && k'_2a'^*_3=k_1a^*_3-k_3a_1a^*_2     \qquad   k'_2a'_3=k_1a_3-k_3a^*_1a_2 \ ,\nonumber\  \label{R123}  \ \eea
with $k'=\sqrt{1-a_2a^*_2}$. This map defines a canonical transformation of the Poisson algebra, which in terms of the angles reads:
$\{\a_i\b_j\}=\delta_{ij}\quad \{\a_i,\a_j\}=0 \quad \{\b_i,\b_j\}=0 $.
We now canonically quantize this theory, {\it i.e.}, by replacing the angles by Hilbert space operators and the Poisson brackets 
by commutators, so that
$[\a,\b]=\zeta \hbar $, where $\zeta$ is a complex parameter.
It can be shown that the quantum  operators corresponding to $ k,a,a^*$, satisfy the commutation relation of the $q$-oscillator algebra
\bea
&& qa^{\dagger}a-q^{-1}a a^{\dagger}=q-q^{-1} \nn\\
&&ka^{\dagger}=q a^{\dagger}k \qquad ka=q^{-1}ak\ ,\nn \\
\eea 
with quantum deformation parameter $q=e^{\zeta \hbar}$ and $k^2=q(1-a^{\dagger}a)$. Upon this quantization, the map  $R_{123}$ becomes a quantum operator  satisfying by construction the quantum tetrahedron equation (\ref{tetraeq}). It has been shown in \cite{Bazhanov-1} that it is possible to construct the matrix elements $\langle n'_1,n'_2,n'_3|R|n_1,n_2,n_3\rangle$  in the basis of the Fock space constructed from the $q$-oscillator algebra. 
The operator $R_{ijk}$ define an automorphism 
of the triplets of tensor products of the $q$-oscillator algebra $ O_q^{\otimes^3}\rightarrow  O_q^{\otimes^3}$, and it also has the property of being 
non-degenerate in $F^{\otimes3}$. This fact, together with the tetrahedron equation (\ref{tetraeq}), implies the 
validity of the standard Yang-Baxter equation (which signals the integrability in two dimensions) . In fact, 
tracing out in the Fock space $F_a$, the following equations are obtained:  
\bea
R_{bc}R_{bd}R_{cd} &=& R_{cd}R_{bd}R_{bc} \label{Yang-Baxter}\\
L_{Vb}L_{Vc}R_{bc} &=& R_{bc}L_{Vc}L_{vb} \label{LLR-RLL} \ .\eea
One affinization of the solution of \cite{Bazhanov-1} has been given in \cite{Bazhanov-2} as follows: consider the layer to layer transfer matrix $T_{mn}(\l,\mu)$ which are related to the $L$-operators by $T_{mn}(\l,\mu)=\prod_{i=1} ^n \prod_{j=m} ^1L_{îj}(\l_i\mu_j)$ and  may be  obtained from an ansatz that solves the local Yang-Baxter equation:
\[ L_{1,2}(u_3,\l_3) =  \left [ \begin{array}{cccc}
1 & 0 & 0          & 0 \\
0 & \l_3  k_3 & a_3^{\dagger} & 0 \\
0 & -q^{-1}\l_3 \mu a_3  &    \mu _3 k_3     & 0  \\
0 & 0 & 0          & -q^{-1}\l_3 \mu_3\  \label{lunif}  \end{array}  \right]\]
where we have chosen the Fock space $F_a$ as $F_3$, for convenience. It has been shown that the solution of the affine TE (when all parameters $\l_i$ are equal to each other) has symmetry $\uqan$, where $n$ is an arbitrary integer and can be considered as the emergent coordinate of the third dimension.

Equations (\ref{Yang-Baxter})(\ref{LLR-RLL}) can be associated either to an integrable two-dimensional statistical system or to a quantum system in $(1+1)$ dimensions. Therefore, the new solution of the Zamolodchikov equation tells us that is possible to break up the $(2+1)$-dimensional lattice system (in a consistent fashion) into $(1+1)$-dimensional ones for each row and column of the lattice or into classical statistical systems in $(2+0)$-dimensions at a fixed time.  
This result relates the one-dimensional Mott point ({\it i.e.}, the critical value of $t/U$ in the one-dimensional analogue of (\ref{Model-Ferm-2d})
and its order parameter)
to its two-dimensional counterpart as follows: on the one hand, in each row or column the $XXZ$ spin system with hamiltonian,
\bea
H_{XXZ}^{1d}=-\sum_{i=1}^{L}(S^x_iS^x _{i+1} + S^y_i S^y_{i+1}
\Delta S^z_iS^z_{i+1})\ +\ H_{b} \label{hxxz2} \eea
where $H_{b}  = \a(S^z_1-S^z_L)$ and $\Delta= (q+q^{-1})/2$ and $\a=(q-q^{-1})/2$, is know to posses the symmetry $U_q(sl(2))\otimes U_q(sl(2)) $ \cite{Sierra-Book}. 
Note that $q=1$ at $\Delta=1$, so that the boundary conditions are irrelevant, and we can choose periodic boundary conditions without breaking the quantum group symmetry. 
The system defined by (\ref{hxxz2}) can be mapped to a one-dimensional nearest neighbors Hubbard (fermionic) system (using a Jordan-Wigner transformation) and may also be bosonized as a Luttinger system, which is a conformal field theory (CFT) \cite{cft} with central charge $c=1$ \cite{Affleck-lecture}\cite{Voight}. It has been shown that this system undergoes a Mott transition at $\Delta=1$ \cite{Shankar}, becoming an insulator. Morever, at the Mott transition (with $q=1$), the quantum group symmetry becomes the ordinary $su(2)$ symmetry (as far as the algebraic and symmetry properties are concerned, we treat $su(2)$ and $sl(2)$ as interchangeable) and  the CFT will have symmetry $\su2\otimes\su2$ (because two CFTs, one for each chirality are needed)  which becomes $su(2)\otimes su(2)$ at long distances. This CFT can be realized by a Wess-Zumino-Witten (WZW) model at level $k=1$, or by two chiral bosons (of opposite chirality) compactified on a circle at the self-dual radius (under $R$-duality) , where the vertex operators, known as currents $J^{\pm}=e^{i\sqrt{2}\phi}$,are well defined under a shift in the zero-mode of the fundamental bosonic field of the WZW model $\phi \rightarrow \phi+2\pi r$ ($r$ is
the compactification radius). The  currents have scaling dimension $(1,0)$ and together with $J^3(z)=i\de \phi$, of the same dimension, satisfy the $\su2$ algebra. The $R$-duality in these systems is well-known (\cite{cft}), for which an exchange of $r$ for $1/r$ leaves the spectrum unchanged 
but the elementary degrees of freedom are exchanged between charges and vortices.  

For $\Delta=1+\epsilon$, the system (\ref{hxxz2}) develops an energy gap in the spectrum $E_g=4\exp(a/\epsilon)$ and  exhibits a charge density wave (CDW) order parameter \cite{Shankar}, defined by $\langle \rho(i) \rangle=1/2[1+(-1)^i P]$, where 
$ P=\langle\psi^\dagger_r (i)\psi_l (i)+ \psi^\dagger_l (i)\psi_r (i)\rangle =  1/\sqrt{\epsilon}\exp{(-a/\sqrt{\epsilon})}$ ($a$ is a constant, 
$i$ labels the lattice site 
and the expetaction value is taken in the ground state). The corresponding low-energy effective theory is given by the Sine-Gordon theory
for the non-chiral effective bosonic field $\phi$, of Lagrangian density:
\bea
L_{SG}=\frac{1}{2}(\de \phi)^2+\beta \cos \sqrt{2} \phi\ . \eea
The last term can be interpreted as having origin in the 
Umkplap processes naturally arising in the lattice fermionic description, in which it is given by $\langle \psi_L\psi_L \psi_R\psi_R \rangle$ (where $\psi_L=\exp(-i\phi_L/\sqrt2)$ and $\psi_R=\exp(i\phi_R/\sqrt2)$, with $\phi=\phi_L+\phi_R$.  As it has been pointed out  in \cite{Nayak}, for $\beta> 0$  (repulsive interactions), the ground state energy is minimized for a constant ground state density $\langle \phi \rangle =\pi/\sqrt{2}$, which in the fermionic representation corresponds to $\langle\psi^\dagger_R\psi_L\rangle = -\langle\psi^\dagger_R\psi_L\rangle =if$, where 
$f$ is a real function that changes sign under a $\pi/2$ rotation around any axis (with appropriate generalizations for the one-dimensional case
\cite{Nayak}) .
Therefore, this one-dimensional system exhibits a $d$-wave density order parameter, meaning that on the ground state the quantity  $\langle \psi^\dagger(k)\psi(k+\pi/(a))\rangle$ breaks individually time reversal, translation invariance by one lattice site and $\pi/2$ rotation (around any axis) symmetries, while preserving the composition of any two of them. (Note that  $\langle\psi^\dagger_R\psi_L\rangle = -\langle\psi^\dagger_R\psi_L\rangle$ yields $P=0$, implying a constant density profile.

As consequence of the projection-like character of the solution of \cite{Bazhanov-1}, 
by consitency,  the system on the lattice must be in a  (two-dimensional) $d$-density wave 
and also will be described by  a CFT with $c=1$ in all phases. In orther to further account 
for the charge neutrality of the spin excitations and the additional lattice symmetries 
\cite{crystal}, the CFT 
should be taken on the orbifold  $S^1/Z_2$ and modded out by the lattice symmetry $D_4$. This CFT has been identified as characterizing the critical point of the six-vertex model or the four-state Potts model \cite{Ginsparg}.

On the other hand, the lattice fermion  system  can be viewed in an alternative way: consider a two-layered system periodic in the 'temporal axis', and trace out over the temporal dimension. The resulting Yang-Baxter equation has symmetry  $\uqa \otimes \uqa $ corresponding to the symmetry of the six-vertex model
\cite{Sierra-Book}, whose phase diagram depends on the Boltzmann weigths at each vertex of the lattice throught the parameter $\Delta_{6v}=(a^2+b^2 -2c^2)/2ab$, where $a=\exp(-\beta E_a)$, $b=\exp(-\beta E_b)$, and $c=\exp(-\beta E_c)$ are the weights at each vertex.
The transfer matrix of the six-vertex model is given by that of the $XXZ$ model through a Wick rotation. At the Mott critical point this implies that 
($\Delta=-\Delta_ {6v } =1$) it  is in the antiferroelectric phase.( Note that the spins are also rotated changing the sign of the spin-wave term \cite{Affleck-lecture})


Now we would like to write down an effective field theory (EFT) for the model on the square lattice at the Mott critical point \cite{Polchinski}. We first choose the effective degrees of freedom, and impose their characteristic symmetries on the theory under construction. In our case, both symmetries are given by the exact solution discussed above. As it was first pointed out by Witten, there is a close relationship between quantum groups, vertex models and Chern-Simons (CS) gauge theories: the expectation values of the Wilson loops can be calculated as statistical sums of Boltzmann weights in suitable defined vertex models, so that the mathematical structure of quantum groups encodes the topology of planar Wilson loops \cite{Witten-Vertex}.
However,  CS theories posses  naturally the symmetry $\uqa$ (with $q=\exp(2\pi i/k)$ where $k$ is the CS coupling constant) \cite{Kogan} \cite{Grinseng}, and not $\uqa \otimes \uqa$ . This mismatch is a consequence of the absence of parity conservation in the CS gauge theories. The simplest CS-type theories that  
preserve parity are the double CS gauge theories (which contain two $u(1)$ chiral gauge fields of opposite chirality, namely right and left)\cite{Carlo-Topics} :
\beq
S_{DCS}=\frac{k}{4\pi} \int d^3x\ a_R \wedge da_R -\frac{k}{4\pi} \int d^3x\  a_L\wedge da_L \label{Double-CS}
\eeq    
where $ a_R$ ($a_L$) denotes the right (left) gauge field.
This theory  is known to be equivalent to the BF theory \cite{Carlo-Supercond} and it can also be written as a mixed CS theory. Here we are using $a \wedge da$ as 
a short-hand notation for the lattice version of the CS coupling $a_\mu K_{\mu,\nu}  a_\nu$ with  $K_{\mu,\nu}=S_{mu}\epsilon_{\mu,\a,\nu}d_\a$, $S_\mu f(x)=f(x+a\epsilon_\mu)$, $S_\mu f(x)=(f(x+a\epsilon_\mu)-f(x))/a$, (with $a$  the lattice spacing). At the Mott transition we have $k=1$, since the coupling $k$ fixes the unit of charge and the statistics of the excitations, and we find consistency with the fact that the effective degrees of 
freedom are density waves of the underlying electron system. Note that the $u(1)$ CS theory can be considered as the broken parity phase of the 
$su(2)$ CS theory, where  the relation to the six-vertex model has been established \cite{Witten-Vertex} \cite{Alekseev}.  
                                                   
We now impose periodic boundary conditions to the EFT, {\it i.e.}, compactify the space domain on a torus. Cutting down the torus along any cycle,
induces the loosing of gauge symmetry on the cycle, so that the gauge fields become boundary dynamical degrees of freedom \cite{Wen}\cite{Witten-2}\cite{Stone} which are free chiral bosons ($c=1$ CFTs), representing charge density waves (CDW) described also by Luttinger systems . In the quantum theory obtained after quantizing these classical bosonic waves, there is a shift in the coupling parameter $k$ that is properly taken into account by the Sugawara construction: $k\rightarrow k+c_v$, where $c_v$ is the dual coxeter number of the symmetry algebra of the gauge group ( $c_v=2$ for $su(2)$). However, the identification of the Mott transition is done at the classical level, implying that the topological order remains given by the relation $ q=\exp(2\pi i /k)$, with no shift in $k$.    

We would now like to show that in the EFT, the emergence of a $d$-wave order considered before is natural.
For this, we focus on the electric-magnetic duality between charges and vortices implicit in the EFT.
Let us consider the charge current $j^\mu(x)$ degrees of freedom in the direct lattice, and
corresponding vortex current $\phi^\mu(X_d) $ in the dual lattice (whose sites are in the center on each paquette of the direct lattice).
We assume that these degrees of freedom can couple, and  
the low-energy action for their interaction is given by a mixed Chern Simons theory
\bea
S_{MCS}=\frac{k}{4\pi}\int d^3x \,a \wedge da -\frac{k}{4 \pi}\int d^3x\, b \wedge db + a_\mu \phi_\mu\   \eea    
where we have introduced two gauge fields $a_\mu$ and $b_\mu $ for the current and vortex degrees of freedom, respectively.
The relevant definitions are $j^\mu=k \hat{K}_{\mu,\nu} a_\nu$, $\phi^\mu=k \hat{K}_{\mu,\nu}b_\nu$,  
$\hat{K}_{\mu,\nu}=\hat{S}_{mu}\epsilon_{\mu,\a,\nu}\hat{d}_\a$, $\hat{S}_\mu f(x)=f(x+a\epsilon_\mu)$, 
$\hat{d}_\mu f(x)=(f(x-a\epsilon_\mu)-f(x))/a$. 
In the dual lattice, the system dual to the original one is a two-dimensional $XXZ$ spin system with 
coupling constant $\Delta^ {-1}$. The degrees of freedom corresponding to these spin variables are vortices. For $\Delta<1$ the spins in the dual system are frozen, there are no spin waves, and the system is in the antiferomagnetic phase. Therefore, its effective action is given by 
a CS theory with punctures\cite{Trugenberger-1}:
\bea
S_{CS}&=&\frac{k}{4\pi}\int d^3x\ a^\mu \ K_{\mu,\nu} a_\lambda + 
\sum^{'}_p \phi^0 \left[ \delta(x_d,y_d)\right . \nn\\
&-& \left . \delta(x_d+1,y)-\delta(x_d,y_d+1) + \delta(x_d+1,y_d+1) \right] \nn\ ,
\label{CS-Vortex} 
\eea
where $a_\lambda$ is a (different) abelian CS field and $\sum^{'}_p$ means that 
the sum is taken over all fundamental domains . Each domain has period $2a$ and contains 
four vortices in antiferromagnetic array.
Therefore, the classical low-lying states reproduce an antiferromagnetic current pattern. At 
the quantum level, Gauss' law selects the physical states from the lattice CS gauge theory 
with punctures. The quantum order of the ground state of this theory 
(staggered flux phase) breaks translation and  parity symmetries by one lattice site and also 
time reversal invariance, but it is, however, 
invariant under the composition of any two symmetry transformations, satisfying the definition of the $d$-density wave order invariance
(\cite{Nayak}). 
Therefore, the EFT analysis shows that the Mott transition for the system defined by (\ref{Model-Ferm-2d}) is of topological ({\it i.e.}, Kosterlitz-Thoulouse type) nature, characterized by the emergence of CS vortices in antiferromagnetic arrays. Note that the resulting theory breaks the $Z_2$ vortex symmetry associated and, therefore, the two-dimensional chirality on each plaquette (\ref{Double-CS}).

Finally, we would like to discuss the behavior of the previously considered EFT away from the Mott critical point. By analogy with the CS theory of the quantum Hall 
effect, we could expect a ground state stable against small doping. In that case, for the simplest Laughlin inverse filling fractions $k=m$ ($m$ odd integer), the ground state is a droplet of incompressible quantum liquid \cite{Laughlin} (however,  
other phases with more exotic quantum orders, like Nematic phases are also possible in other regimes) and is stable under small perturbations away from
the center of a given plateu in the conductivity.
In the Mott system, we have already assumed that the dynamically generated vortices act as external statistical fields for the new electrons 
injected in the system by doping (this can be considered as an extension of the $R$-duality).At the self-dual point, statistical magnetic fields can be interchanged with statistical electric fields (on a torus). After imposing the lattice symmetries, the low-lying effective Hamiltonian for the injected electrons (in first quantization) is: 


\bea
H=\sum_i [-\hbar^2\frac{1}{2m}(\frac{\de^2}{\de x_i ^2} +\frac{\de^2}{\de y_i^2})+\lambda _i(x_i^2- y_i^2)] \ ,\label{effective-force} \eea
where $\lambda _i$ can take the values $\pm \lambda$
Therefore, the electric potential changes sign in $x=\pm y$, producing domain walls between regions with different electron densities. Similar results can be obtained using the $W_4$ symmetry, which is related to the relevant perturbations of the Ashkin-Teller and six-vertex models away from the critical point \cite{Bottesi-Zemba}\cite{Gaite}.

One consequence of having discussed the EFT is that, {\it a-posteriori}, the behavior of the electrons can be more easily understood. 
It can be shown that the interaction term in the hamiltonian (\ref{Model-Ferm-2d}) in the the continuum limit contains a
chemical potential term of the form $-\mu\ \rho$, with $\mu=\Delta$ , which ensures the half-filling condition. Therefore, changing  the chemical potencial by doping in $\delta \mu$ modifies the hamiltonian (in the spin representation (\ref{hxxz}) by $H(\Delta)\rightarrow H(\Delta)+\delta \mu \sum_{\langle i j\rangle} S^z_i S^z_j $. At $\Delta=1$,the dynamics of the electron system is given by the double CS theory (\ref{Double-CS}), whose hamiltonian can be defined as the temporal component of the stress-energy tensor $H_{cs}=T_{00}$, where $T_{\mu\nu}=\delta S_{cs}/\delta g_{\mu\nu}$  and where $g_{\mu\nu}$ is the metric tensor. However, the CS action is topological and, therefore, independent of the metric and $H_{cs}=0$
for each chiral component leading to $H(\Delta=1)=0$. This means that doping the system away from the critical point, the dynamics is controlled by an effective Ising hamiltonian.

To sum up, our study of the EFT for the model (\ref{Model-Ferm-2d}) shows that the Mott transition is of topological ({\it i.e}, Kosterlitz-Thoulouse type) nature, characterized by the emergence of CS vortices in antiferromagnetic arrays. The symmetry $\su2\otimes\su2$ of the critical point implies that electric and magnetic vortices can be intercanged, providing  effective atractive and repulsive forces.  Doping the system  produces domain walls, signaling a quantum gas-liquid phase coexistence, which belongs to the Ising universality class.



 \def\RMP{{\it Rev. Mod. Phys.\ }}

 \def\PRL{{\it Phys. Rev. Lett.\ }}

 \def\PL{{\it Phys. Lett.}}

 \def\PR{{\it Phys. Rev.  \ }}
 \def\NP{{\it Nuclear. Phys.}}
 \def\PRB{{\it Phys. Rev. B  \ }}
 \def\IJMP{{\it Int. J. Mod. Phys.}}
  
\end{document}